\documentclass[twocolumn]{aastex631}

\newcommand\asloth{\textsc{a-sloth}}
\newcommand{\Msun}{\,\ensuremath{\mathrm{M}_\odot}}

\begin{document}

\title{Unveiling the contribution of Pop III stars in primeval galaxies at redshift $\geq 6$}



\correspondingauthor{Muhammad A. Latif}
\email{latifne@gmail.com}

\author[0000-0003-3518-0235]{Shafqat Riaz}
\affiliation{Center for Field Theory and Particle Physics and Department of Physics, 
Fudan University, 200438 Shanghai, China}

\author[0000-0001-6742-8843]{Tilman Hartwig}
\affiliation{Institute for Physics of Intelligence, School of Science, The University of Tokyo, Bunkyo, Tokyo 113-0033, Japan}
\affiliation{Department of Physics, School of Science, The University of Tokyo, Bunkyo, Tokyo 113-0033, Japan}
\affiliation{Kavli Institute for the Physics and Mathematics of the Universe (WPI), The University of Tokyo Institutes for Advanced Study, The University of Tokyo, Kashiwa, Chiba 277-8583, Japan}

\author[0000-0003-2480-0988]{Muhammad A. Latif}
\affiliation{Physics Department, College of Science, United Arab Emirates University, PO Box 15551, Al-Ain, UAE}


\begin{abstract}
Detection of the first stars has remained elusive so-far but their presence may soon be unveiled by upcoming JWST observations. Previous studies have not investigated the entire possible range of halo masses and redshifts which may help in their detection.  Motivated by the prospects of detecting galaxies up to $z\sim 20$ in JWST early data release,  we quantify the contribution of Pop III stars to high-redshift galaxies from $6 \leq z \leq 30$ by employing the semi-analytical model \asloth, which self-consistently models the formation of Pop III and Pop II stars along with their feedback. Our results suggest that the contribution of Pop III stars is the highest in low-mass halos of $\rm 10^7-10^9~\Msun$. While high-mass halos $\rm \geq 10^{10}~\Msun$ contain less than 1\% Pop III stars, they host galaxies with stellar masses of $\rm 10^9~\Msun$ as early as $z \sim 30$. Interestingly, the apparent magnitude of Pop~III populations gets brighter towards higher redshift due to the higher stellar masses, but Pop~III-dominated galaxies are too faint to be directly detected with JWST. Our results predict JWST can detect galaxies up to $z\sim 30$, which may help in constraining the IMF of Pop III stars and will guide observers to discern the contribution of Pop~III stars to high-redshift galaxies.

\end{abstract}


\section{Introduction} \label{sec:intro}
Tremendous progress on the observational frontier has enabled astronomers to detect galaxies  up to the cosmic dawn during the past two decades. About thousand galaxies have been detected detected at z $>$6  \citep{Bowens16,Oesch16,Harikane22,schaerer22,Finkel22} with candidates up to z$\sim$ 20 being revealed in James Webb space telescope (JWST) early data release \citep{carnall22,castellano22,naidu22,Yan22,Adam22}, which may just be the tip of the iceberg \citep{dayal18}. One of the primary goals of JWST is to unveil primeval galaxies that contain Pop III stars and revolutionize our understanding of the high-redshift universe. In fact, the commissioning of JWST has shown that unprecedented sensitivity of NIRCam can detect objects with a flux of $\sim 10$\,nJy (equivalent to  apparent  magnitude of $\sim$ $29$) at SNR$=10$ for a standard exposure time of 10\,ks \citep{rigby22}. With longer exposure times and gravitational lensing, JWST may discover even more and fainter galaxies at redshift $>10$. Therefore, it is very timely to make predictions about the contribution and presence of Pop III stars. Such work will help in guiding forthcoming JWST observations.

Pop III stars are expected to form in pristine minihalos of a few times $\rm 10^6 \Msun$ at z $\geq 10$ \citep{skinner20,schauer21}. They ushered the universe out of cosmic dark ages, initiated the process of re-ionization and shaped the formation of high-redshift galaxies via their feedback. Depending on their mass, they are expected to have short lifetimes, may go off as a supernovae (SNe) and enrich the surrounding ISM with metals \citep{Heger02}. In the aftermath of Pop III SNe, second generation stars known as Pop II stars form from metal-enriched gas with metallicity as low as $\rm \geq 10^{-5}~Z_{\odot}$ \citep{Schneider03,Omukai05}.  Recent, numerical simulations including UV radiative feedback from stars suggest Pop III characteristic masses of a few ten of $\rm \Msun$ \citep{Clark11, Stacy16,Sugi20,Latif22} substantially lower than previously thought \citep{Bromm02,Abel02,Yoshida08}. However, direct observations are required to constrain their mass spectrum, which might be achieved with upcoming observations of high-redshift galaxies with JWST.

In fact, the star formation rate density (SFRD) of Pop~III stars dominates at $z \geq 15$ \citep{hartwig22} suggesting their significant role in shaping high-redshift galaxies and the necessity of taking into account the contribution of Pop III stars in  modeling their SEDs.  \citet{zackrisson11,zackrisson17} investigated the spectral evolution of first galaxies finding that Pop III galaxies  with stellar masses as low as $\rm 10^5 \Msun$ can be detected at z $\sim$ 10 and discussed various observational strategies. Renaissance simulations have examined the properties of high-redshift galaxies such as their stellar masses, SFRs, UV luminosity functions and escape fractions of ionizing radiation \citep{oShea15,xu16}. Their results suggest  that large fractions of Pop III stars may remain elusive to direct detection at z=15 \citep{barrow18}. \citet{jaacks19} studied the legacy of Pop III star formation and found that the Pop~III contribution to SFRDs significantly increases beyond z$\sim$15 up to 50\% while their contribution to ionizing emissivity is about 60 \%.  Recently, \citet{katz22} simulated  a halo of $\rm 3 \times 10^8 \Msun$ at z=10 to investigate the possibility of Pop~III detection with JWST and found that key signatures of Pop III stars fade away quickly due to their short lifetimes. These studies could not investigate the entire range of possible halo masses and redshifts due to numerical limitations, but they indicate that Pop III stars might be detectable at high redshift.

Motivated by the prospects of detecting galaxies with JWST up to $\rm z\sim 20$ \citep{Yan22}, in this letter, we perform a comprehensive study which self-consistently models the formation of both Pop III \& Pop II stars along with their chemical, mechanical and radiative feedback for a statistical sample of high-redshift galaxies.  We simulate here a wide range of galaxies forming in different halo masses at $z=6-30$ because of the expected dominance of Pop III stars in this era and report their properties, such as masses and luminosities for Pop~III and Pop~II stars. These results will help to identify the possible contributions of Pop~III stars in the upcoming data of JWST.

\section{Methodology} \label{sec:methods}
We use the semi-analytical model \asloth\ to simulate the formation of the first galaxies \citep{hartwig22,magg22}. The model is based on dark matter merger trees, which are generated with the Extended Press-Schechter formalism \citep[EPS,][]{PressSchechter,Bond1991}. Given a halo mass and final redshift, the code first generates a dark matter merger tree backwards in time and then simulates the baryonic physics and feedback forward in time. Stars form once a halo is above the critical mass for efficient gas cooling \citep{schauer21}, which includes the baryonic streaming velocity and a Lyman-Werner background, following \citet{hartwig22}. \asloth\ includes chemical, radiative, and mechanical feedback from stars and different types of SNe and distinguishes between Pop~III and Pop~II star formation based on the ISM composition \citep{chiaki17}, which results in an effective threshold metalliticty of around $10^{-5}\,Z_\odot$.

We sample individual stars based on pre-defined initial mass functions (IMFs) for Pop~III and Pop~II stars. This allows us to trace the lifetime and feedback of stars and their supernova explosions accurately in time. Moreover, we can precisely determine the surviving stellar mass at any redshift based on their lifetimes. Hence, our model does not rely on analytical star formation histories, or assumes a single stellar population. Instead, we model the formation of individual stars in high-redshift galaxies self-consistently.

The model is calibrated based on six observables, such as the ionization history and the cosmic SFRD, which guarantees reliable predictions up to high redshifts. For Pop~II stars, we assume a Kroupa IMF in the mass range 0.1-100\Msun while for Pop~III stars we employ a logarithmically flat IMF in the mass range $5-210\Msun$, which best reproduces observations \citep{hartwig22}. However, the lower-mass end of the Pop~III IMF is poorly constrained and for this research we allow Pop~III stars to form down to $3\Msun$. This does not change the global properties of the model, but it allows us to show more fine-grained results due to their slightly longer lifetimes.

\asloth\ resolves minihalos with $M_h \geq 10^6\Msun$ at the highest redshifts. While this high mass resolution is excellent to follow the physics accurately, such small galaxies might suffer from stochastic sampling effects because they only contain a handful of stars. Therefore, we resample each galaxy several times with different random seeds and report the median value and the central $68\%$ percentile to quantify the cosmic variance.

\begin{figure*}[htb]
    \centering
    \includegraphics[width=1.0\textwidth]{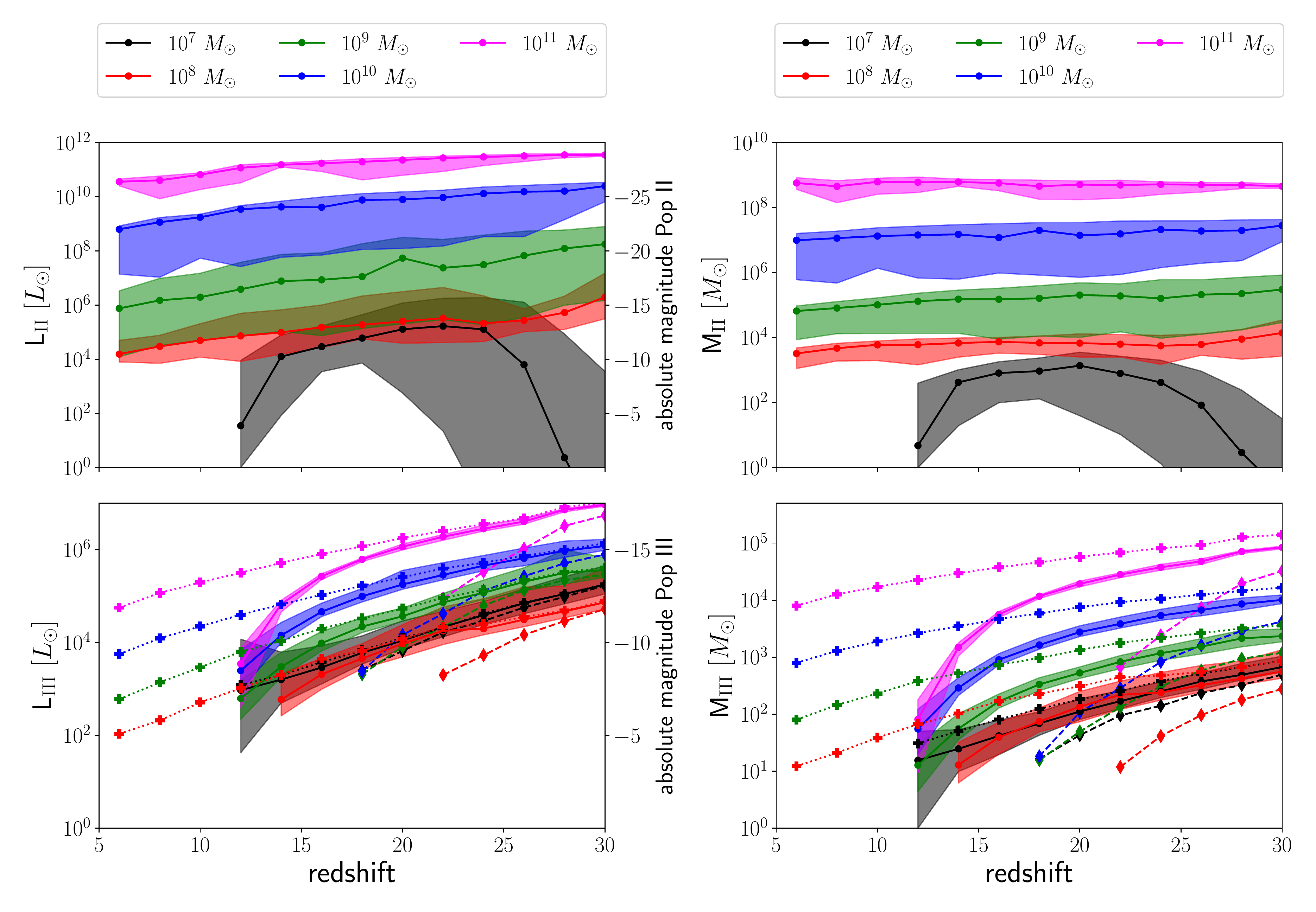}
    \caption{Luminosities, absolute magnitudes (left panels) and stellar masses (right panels) of both Pop~II \& Pop~III stars surviving until the respective redshift are shown in the top and the bottom panels, respectively. In the bottom panels plus signs (dotted lines) and diamonds (dashed lines) are for M$_{\mathrm{min}} = 0.8$ and $5~$M$_{\mathrm{\odot}}$, respectively. In each panel, the filled color region represents the central 68 percent cosmic variance.}
    \label{fig1}
\end{figure*}

\section{Results\label{sec:results}}
We have simulated a large number of high-redshift galaxies using \asloth, which allows to self-consistently model the formation of Pop III and Pop II stars. Moreover, we have explored a variety of halos  with masses ranging from $\rm 10^7-10^{11}~\Msun$ from  $\rm z=30$ down to $\rm z=6$. To mimic cosmic stochasticity, each combination of redshift and halo mass is simulated about 100 times. This comprehensive study enables us to estimate the properties of high-redshift galaxies such as stellar masses, star formation rates, metallicities, luminosities and to quantify the relative contribution of Pop III stars.

The average Pop III stellar mass varies from $\rm 10-10^5 ~\Msun$ for halos of $\rm 10^7-10^{11}~\Msun$ and increases with redshift, see Fig.~\ref{fig1}. At $z \leq 10$, Pop III stars form only in $<25\%$ of simulated halos due to metal pollution. The Pop II stellar mass varies from $\rm 100-10^9 ~\Msun$ for $\rm 10^7-10^{11}~\Msun$ halos. Overall, the Pop II stellar mass does not significantly change in similar mass halos from $\rm z=30 - 6$ as shown in Fig.~\ref{fig1}. Statistical variations in Pop III stellar mass from halo to halo are within a factor of a few and prominent in low-mass halos as they are more prone to stellar feedback and effects from random sampling.

We have selected here a fudicial value of  $\rm M_{min} =3 \Msun\ $ for the lower cutoff mass which is the logarithmic mean of the possible range of minimum Pop III stellar masses between $\rm 0.8-10 \Msun\ $ \citep{hartwig22} and therefore the most representative value. We also investigated the impact of lower cutoff masses on the survival of Pop III stars. Our findings suggest that for a cutoff mass of $\rm  5 \Msun$, Pop III stars stop contributing already at z=18 but for lower cutoff mass of $\rm 0.8 \Msun$ they can survive down to z $\sim$ 6. Overall, for the lowest cutoff mass, the Pop III stellar mass is a factor of few higher than our fiducial case at all redshifts.

Typical luminosity of Pop III stellar populations varies from $\rm 10^3-10^7~L_{\odot}$ while for Pop II it ranges from $\rm 10^3-10^{12}~L_{\odot}$ as depicted in Fig. \ref{fig1}.  Both Pop III \& Pop II luminosities increase with redshift and are highest at $\rm z\sim 30$. For high-mass halos, the Pop III luminosity is a few orders of magnitude smaller than Pop II, but for halos with masses $\rm 10^7-10^8~\Msun$, the Pop III luminosity is roughly comparable to Pop II.  These results suggest that some massive galaxies with stellar mass of $\rm 10^{9}~\Msun$ at $\rm z>15$ have luminosities of $\rm 10^{12}~L_{\odot}$ and are as bright as their counterparts at $\rm z \sim 10$ \citep{naidu22}.

\begin{figure}[htb]
    \centering
    \includegraphics[width=0.47\textwidth]{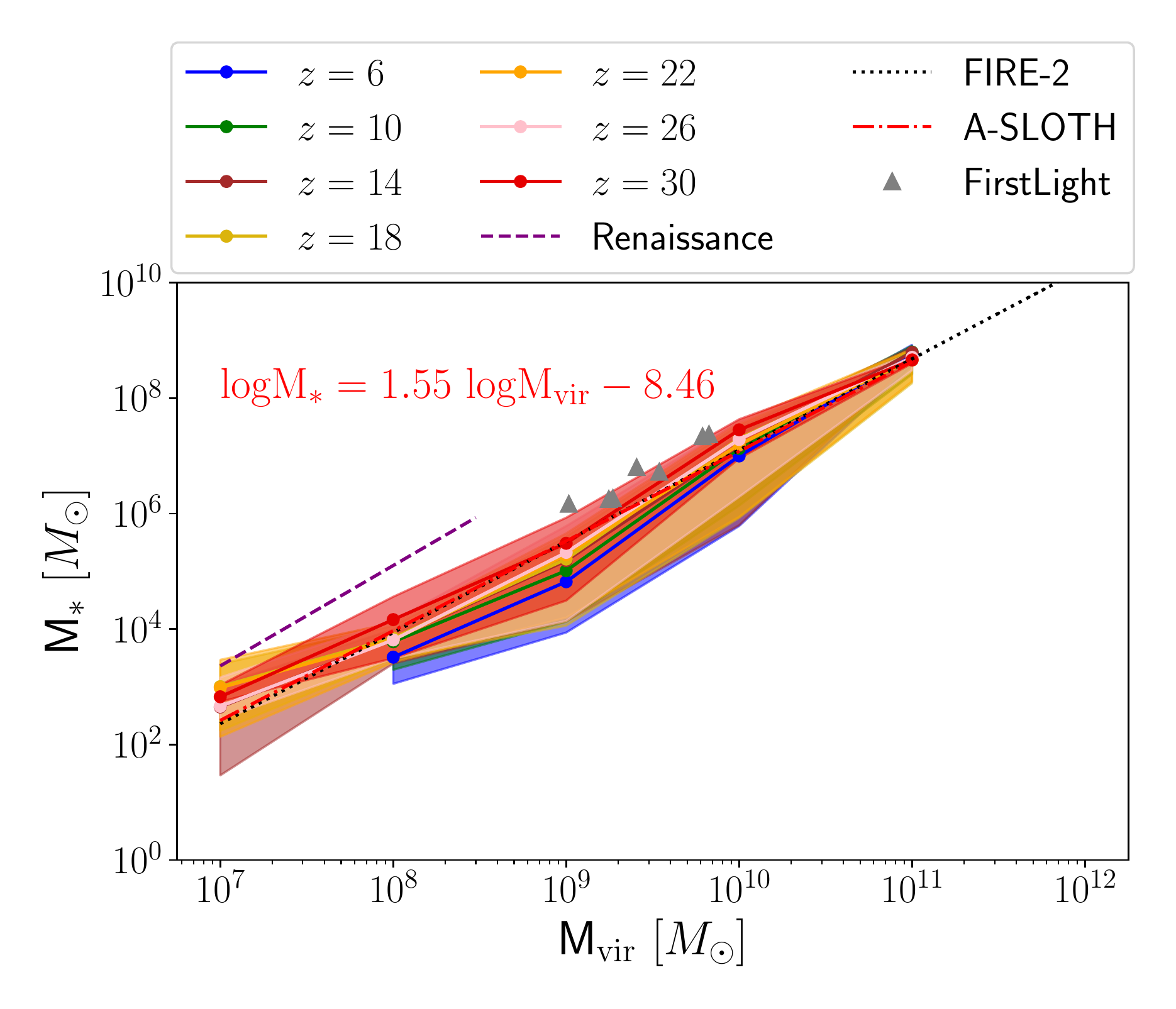}
    \caption{Stellar mass (M$_{*}$ = M$_{\rm II}$ + M$_{\rm III}$) vs halo virial mass (M$_{\mathrm{vir}}$) for redshifts between $6 - 30$. Stellar mass consists of both Pop~II \& Pop~III stars that have survived until the respective redshift. The solid lines represent the median derived from \asloth\ with different random seeds and the colored region shows the central 68 percentile variance. The red dash-dotted line and formula show a linear fit to our data. It is in very good agreement with the results from the FIRE-2 simulation \citep{ma18}. The gray triangles represent data from the First Light simulation at $z = 9.6$ \citep{Ceverino17} and the dashed purple line depicts the fit taken from the Renaissance simulation \citep{oShea15}.}
    \label{fig2}
\end{figure}
Our estimates for stellar mass vs halo mass for the entire sample of galaxies from $\rm z=30-6$ are shown in Fig. \ref{fig2}. Total stellar mass varies from $\rm \sim 10^2 ~to~ 10^9 \Msun$ for halo masses of $\rm 10^7-10^{11}~\Msun$ and monotonically increases with halo mass. Scatter in the plot is due to the statistical variations in the merger trees and IMF sampling. We find no statistically significant change in the stellar mass to halo mass relation in the redshift range $6 \leq z \leq 30$, i.e., the results at all redshifts lie within their uncertainty range. Overall, our results are in good agreement with previous studies \citep{oShea15,Ceverino17,ma18,jaacks19,pallottini22}. However, the Renaissance simulation predicts more stars at a given halo mass, which according to \citet{Ceverino17} is due to the inefficient feedback.

\begin{figure*}[htb]
    \centering
    \includegraphics[width=1.0\textwidth]{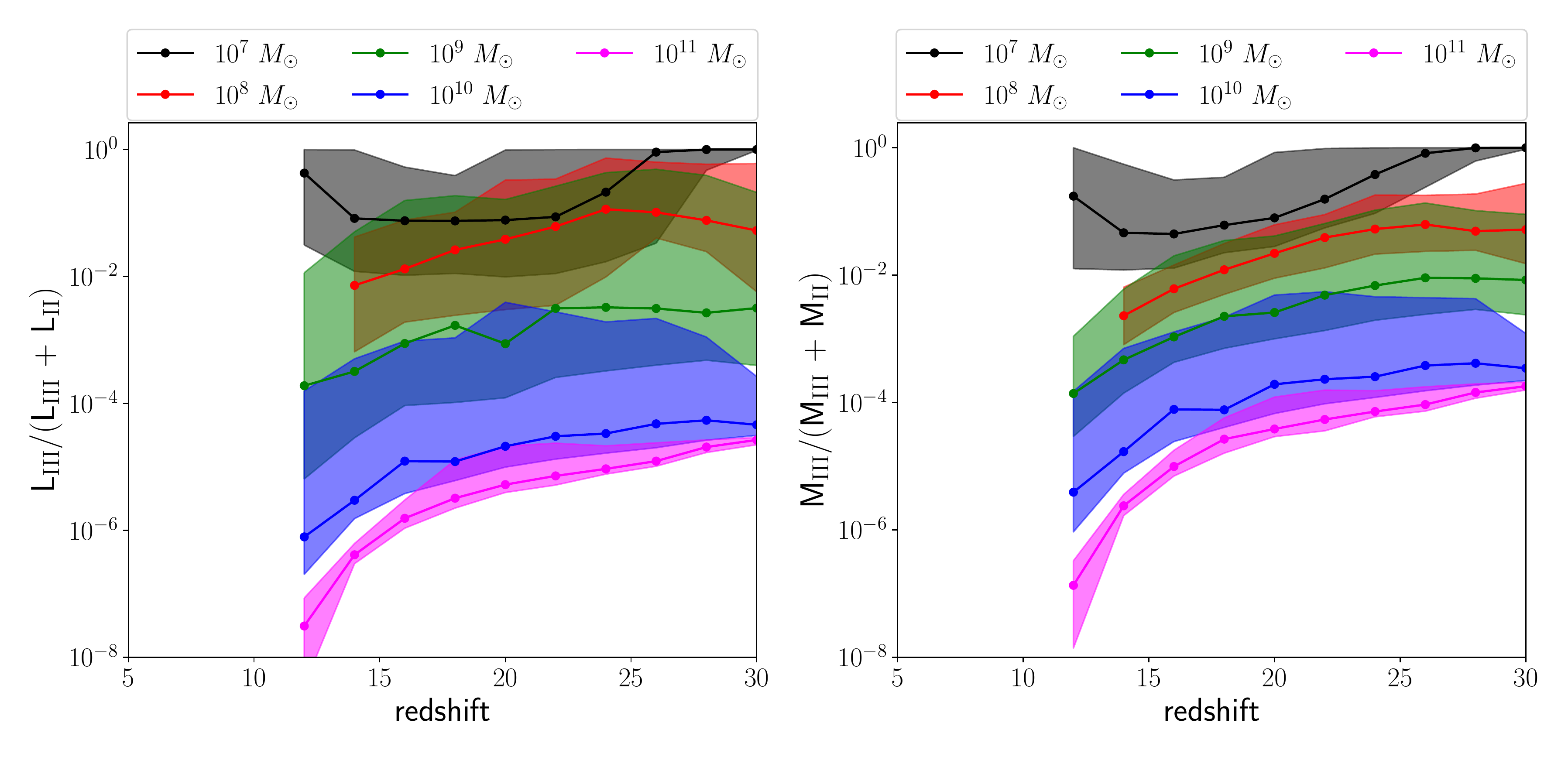}
    \caption{The fractions of Pop~III luminosity (left panel) and stellar mass (right panel) for our simulated galaxies are shown as a function of redshift. The halo mass varies from $10^{7} - 10^{11}~M_{\odot}$ and the colored region represents the central 68 percentile variance.}
    \label{fig3}
\end{figure*}
To further elucidate the contribution of Pop III stars to high-redshift galaxies, we show the ratios of Pop III to total stellar mass and Pop III  to total luminosity in Fig. \ref{fig3}. It shows that contribution of Pop III stars is close to unity at $\rm z>25$ and highest in low-mass halos at all times from z=30 down to z=10. This contribution drops to below 1\% for halos  of $\rm \geq 10^{10}~\Msun$. Furthermore, statistical variations in the Pop III contribution to high-redshift galaxies are two orders of magnitude. Our findings suggest that low-mass galaxies forming at $\rm z\geq 12$ are the best targets to find Pop III stars. The contribution of Pop III stars in high-mass galaxies is much lower than Pop II stars which may pose a challenge to identify them. 

\begin{figure*}[htb]
    \centering
    \includegraphics[width=1.0\textwidth]{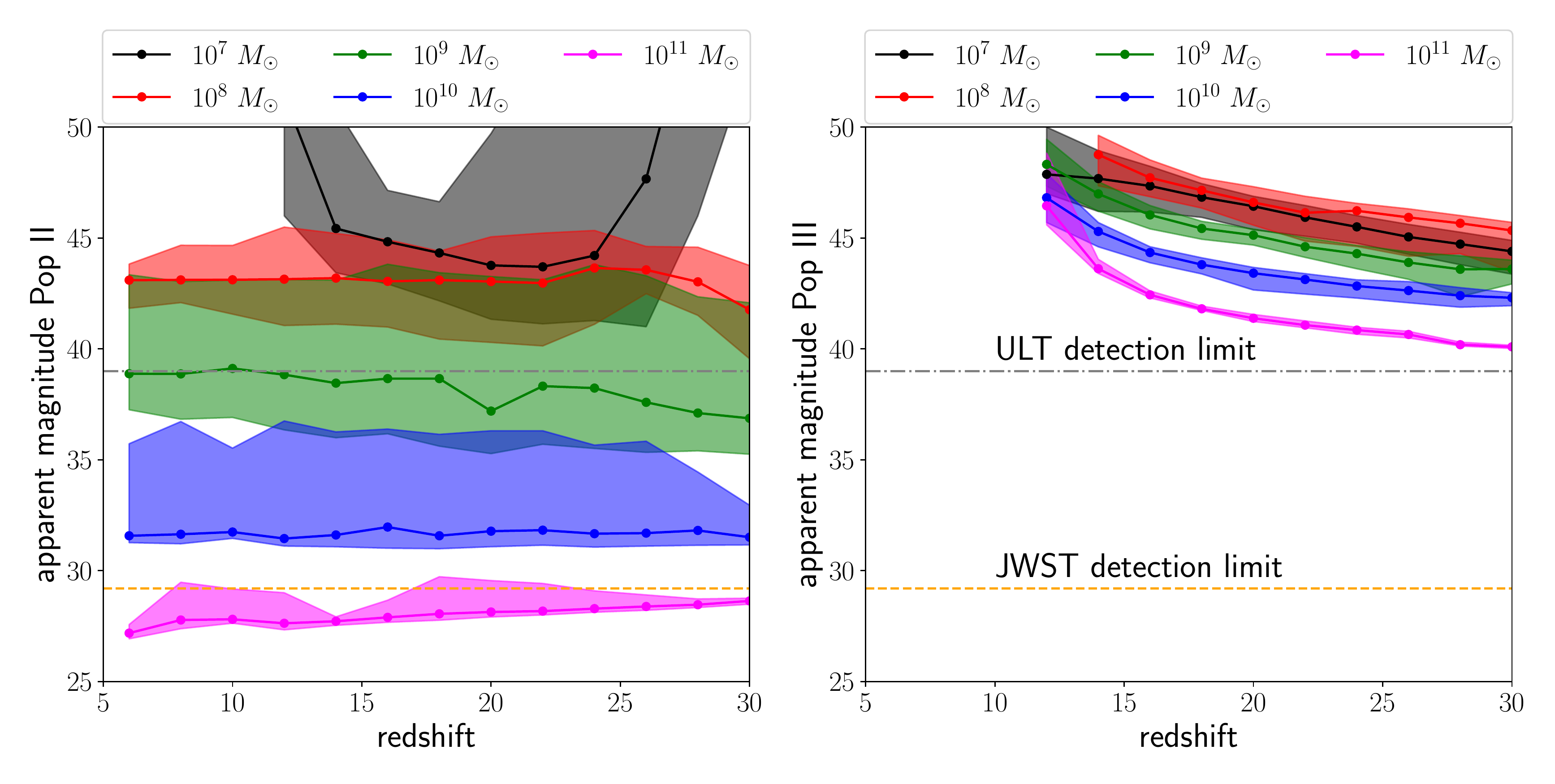}
    \caption{The apparent magnitude of Pop~II (left panel) and Pop~III (right panel) stars that survived until given redshift as a function of redshift. The halo mass ranges from $10^{7} - 10^{11}~M_{\odot}$. The orange dashed line in both panels denotes the JWST  apparent magnitude limit of 29.2 (SNR=10 for an exposure time of 10\,ks in F277W, \citealt{rigby22}). The colored region quantifies the central 68 percentile variance. In the both panels, the gray line at the apparent magnitude of 39 represents the detection limit of Moon-based near-infrared mission, The Ultimately Large Telescope (ULT), proposed in~\citet{schauer20}. }
    \label{fig4}
\end{figure*}
To compare our results with observations, we estimated the bolometric apparent magnitudes of our galaxies and their statistical variations, which are shown in Fig. \ref{fig4}. To calculate the bolometric apparent magnitude, we used $m = -26.83 - 2.5\log\left(F/F_{\odot}\right),$ where $F_{\odot}$ is the Solar flux, and $F$ is the flux of our simulated galaxy, which is estimated using the luminosity distance relation for a given redshift. We compute the luminosities of Pop III \& Pop II stars separately. For Pop III stars, we use the fitting function given in Eq 3 of  \cite{Winhorst18} while for Pop II stars, we use a standard luminosity-mass relation. We also show the expected apparent magnitude of Pop III and Pop II stars separately that enables us to quantify their contributions and compare them with the detection limit of JWST. The apparent magnitude of Pop III stars varies from 40-50, increases with redshift, and is brightest for a $\rm 10^{11}\,\Msun$ halo at $\rm z\sim 30$, much fainter than the JWST detection limit of 29.2. The range of apparent magnitude for Pop II stars varies from 27-50, which is much larger than for Pop III stars, but only the most massive galaxy will be visible to JWST. In fact, such a galaxy can be detected as early as z=30.

\begin{figure*}[htb]
    \centering
    \includegraphics[width = 0.70\textwidth]{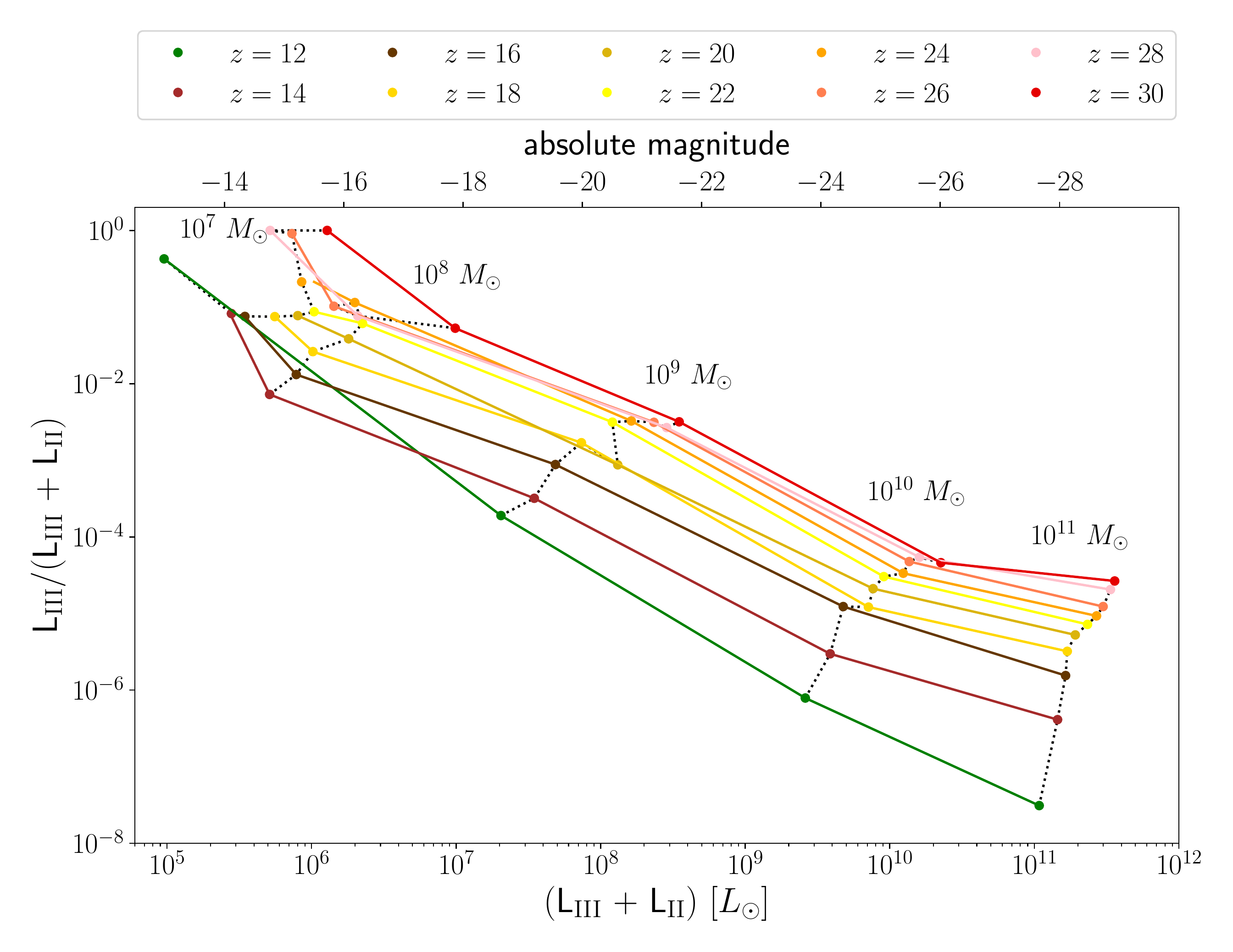}
    \caption{The luminosity fraction of Pop~III stars as a function of the total luminosity and absolute magnitude for different halos as depicted above each curve. We consider only the surviving Pop~II and Pop~III stars at the final redshift. The color of the dot reflects the final redshift and the colored solid line represents the trend in luminosity fraction with increasing halo mass at a given redshift. }
    
    \label{fig5}
\end{figure*}
We further show the fraction of Pop III luminosity against the total luminosity/AB magnitude for the entire sample of simulated galaxies in Fig. \ref{fig5}. This figure provides a convenient way to estimate the relative contribution of Pop~III stars to the total luminosity of newly detected galaxies at high redshift. It is found that faint galaxies with AB magnitude below -20 are the best candidates for finding Pop III stars across all redshifts but are well below the detection limits of JWST. The brightest galaxies contain less than 1\% Pop III stars but their AB magnitudes are within the range of JWST even at z=30. Based on the halo mass function of \cite{Warren06}, the expected number density of such galaxies is $\rm \leq 1$ per Gpc$^3$ at z=26. Therefore, these galaxies are expected to be rare at earlier times. Nevertheless,  we expect such galaxies can be unveiled in upcoming wide survey JWST observations.

\section{Discussion and Conclusions}
We have simulated a large ensemble of high-redshift galaxies using the semi-analytical model \asloth, which has been calibrated against six independent observables and simultaneously models both Pop~III \& Pop~II stellar populations along with their feedback. This unique sample of galaxies allowed us to study the statistical variations among the properties of high-redshift galaxies, such as stellar masses, luminosities, star formation rates, fraction and contribution of Pop III stars to their host galaxies. Our results suggest that best candidates to search for Pop III stars are low-mass galaxies from $10 \leq z \leq 30$ which are challenging to be detected with JWST and the contribution of Pop III stars decreases to less than 1 \% in massive galaxies. We further predict that JWST can detect galaxies up to $z \sim 30$ as their AB magnitudes lie within its range. These findings may guide observers in planning their observations and also help to improve spectral modeling of high-redshift galaxies.

We consider the impact of both baryonic streaming motions and LW radiation based on \cite{schauer21}, which increase the halo threshold mass above $\rm 10^6 \Msun$. Therefore, Pop III stars cannot form in halos with masses lower than this at $z<20$. Pop III stars can still form in halos of a few times $\rm 10^6 \Msun\ $ at $z>20$. They have typical masses of about $10^3\Msun$ but their feedback limits the formation of Pop II stars in the host halos. Interestingly, these halos hosting very young Pop III star may be directly detected with ULT in future.

We also investigated the role of the Pop III IMF on our findings by varying its slope from logarithmically flat to Salpeter and found that it has negligible impact on the number of Pop III survivors and total stellar mass. Furthermore,  we found that the low  cutoff mass of Pop III IMF influences the number of Pop III survivors as well as their masses.  We find that Pop III stars can only survive to $z \sim 6$ if their low cutoff mass is $\rm < 3 ~ \Msun$ otherwise they die on relatively short timescales and may not survive to such redshifts. Higher cutoff mass ($\rm 5 \Msun$)  decreases the number of Pop III survivors. These results suggest that finding Pop III survivors at $z\leq 10$  may help in constraining the lower mass end of the Pop III IMF.

Our results are in agreement with previous works, which simulated only a limited number of high-redshift galaxies (see Fig.~\ref{fig2}). In addition, \citet{barrow18} find similar Pop III stellar masses at a given halo mass in the Renaissance simulation. They also report the fraction of Pop~III stellar mass to be in the range $10^{-6}-0.3$ for halos of $10^7-10^{10}\Msun$, similar to our results. Recently, \citet{Yan22} report the discovery of three galaxy candidates with a photometric redshift of $z \sim 20$ with JWST. These galaxies have stellar masses of $\sim 10^8\Msun$. Based on our results, we can estimate the dark matter halo mass of such galaxies to be $10^{10}-10^{11}\Msun$ (Fig.~\ref{fig2}). This allows us to estimate the contribution of Pop~III stars to the bolometric luminosities of such objects to be only $\lesssim 10^{-3}$ (Fig.~\ref{fig3}). The contribution of Pop~III stars to the luminosity of such objects is hence negligible.

In this work, we employed the stochastic feedback model of \asloth ~which is based on the EPS merger trees. Although this stochastic feedback is sufficiently accurate \citep{hartwig22}, future studies can improve by using \asloth's spatial feedback model based on merger trees extracted from N-body simulations to obtain more realistic results. If we were to perform 3D cosmological simulations, which are prohibitively expensive for such a large sample of galaxies, pockets of metal free gas may exist down to lower redshifts due to inhomogeneous mixing of metals. For example, \cite{Liu20} found from cosmological simulations that such pockets of metal free  exist even down to z $\sim$ 4 in massive halos of $\rm \geq  10^{9} ~\Msun$.  Under these conditions, some Pop III stars may form at lower redshifts.

\begin{acknowledgments}
TH acknowledges funding from JSPS KAKENHI Grant Numbers 19K23437 and 20K14464. MAL thanks UAEU for funding via SURE Plus 3835 and UPAR grant No. 31S390. 
\end{acknowledgments}


\software{\textsc{a-sloth} \citep{hartwig22,magg22}, python \citep{python09}, numpy \citep{harris20}, scipy \citep{virtanen20}, matplotlib \citep{hunter07}, astropy \citep{price2018astropy}.}


\begin{thebibliography}{}
\expandafter\ifx\csname natexlab\endcsname\relax\def\natexlab#1{#1}\fi
\providecommand{\url}[1]{\href{#1}{#1}}
\providecommand{\dodoi}[1]{doi:~\href{http://doi.org/#1}{\nolinkurl{#1}}}
\providecommand{\doeprint}[1]{\href{http://ascl.net/#1}{\nolinkurl{http://ascl.net/#1}}}
\providecommand{\doarXiv}[1]{\href{https://arxiv.org/abs/#1}{\nolinkurl{https://arxiv.org/abs/#1}}}

\bibitem[{{Abel} {et~al.}(2002){Abel}, {Bryan}, \& {Norman}}]{Abel02}
{Abel}, T., {Bryan}, G.~L., \& {Norman}, M.~L. 2002, Science, 295, 93,
  \dodoi{10.1126/science.295.5552.93}

\bibitem[{{Adams} {et~al.}(2022){Adams}, {Conselice}, {Ferreira}, {Austin},
  {Trussler}, {Juod{\v{z}}balis}, {Wilkins}, {Caruana}, {Dayal}, {Verma}, \&
  {Vijayan}}]{Adam22}
{Adams}, N.~J., {Conselice}, C.~J., {Ferreira}, L., {et~al.} 2022, arXiv
  e-prints, arXiv:2207.11217.
\newblock \doarXiv{2207.11217}

\bibitem[{{Barrow} {et~al.}(2018){Barrow}, {Wise}, {Aykutalp}, {O'Shea},
  {Norman}, \& {Xu}}]{barrow18}
{Barrow}, K. S.~S., {Wise}, J.~H., {Aykutalp}, A., {et~al.} 2018, \mnras, 474,
  2617, \dodoi{10.1093/mnras/stx2973}

\bibitem[{{Bond} {et~al.}(1991){Bond}, {Cole}, {Efstathiou}, \&
  {Kaiser}}]{Bond1991}
{Bond}, J.~R., {Cole}, S., {Efstathiou}, G., \& {Kaiser}, N. 1991, \apj, 379,
  440, \dodoi{10.1086/170520}

\bibitem[{{Bouwens} {et~al.}(2016){Bouwens}, {Oesch}, {Labb{\'e}},
  {Illingworth}, {Fazio}, {Coe}, {Holwerda}, {Smit}, {Stefanon}, {van Dokkum},
  {Trenti}, {Ashby}, {Huang}, {Spitler}, {Straatman}, {Bradley}, \&
  {Magee}}]{Bowens16}
{Bouwens}, R.~J., {Oesch}, P.~A., {Labb{\'e}}, I., {et~al.} 2016, \apj, 830,
  67, \dodoi{10.3847/0004-637X/830/2/67}

\bibitem[{{Bromm} {et~al.}(2002){Bromm}, {Coppi}, \& {Larson}}]{Bromm02}
{Bromm}, V., {Coppi}, P.~S., \& {Larson}, R.~B. 2002, \apj, 564, 23,
  \dodoi{10.1086/323947}

\bibitem[{{Carnall} {et~al.}(2022){Carnall}, {Begley}, {McLeod}, {Hamadouche},
  {Donnan}, {McLure}, {Dunlop}, {Bondestam}, {Cullen}, {Jewell}, \&
  {Pollock}}]{carnall22}
{Carnall}, A.~C., {Begley}, R., {McLeod}, D.~J., {et~al.} 2022, arXiv e-prints,
  arXiv:2207.08778.
\newblock \doarXiv{2207.08778}

\bibitem[{{Castellano} {et~al.}(2022){Castellano}, {Fontana}, {Treu},
  {Santini}, {Merlin}, {Leethochawalit}, {Trenti}, {Mestric}, {Vanzella},
  {Bonchi}, {Belfiori}, {Nonino}, {Paris}, {Polenta}, {Roberts-Borsani},
  {Boyett}, {Calabro}, {Glazebrook}, {Grillo}, {Mascia}, {Mason}, {Mercurio},
  {Morishita}, {Nanayakkara}, {Pentericci}, {Rosati}, {Vulcani}, {Wang}, \&
  {Yang}}]{castellano22}
{Castellano}, M., {Fontana}, A., {Treu}, T., {et~al.} 2022, arXiv e-prints,
  arXiv:2207.09436.
\newblock \doarXiv{2207.09436}

\bibitem[{{Ceverino} {et~al.}(2017){Ceverino}, {Glover}, \&
  {Klessen}}]{Ceverino17}
{Ceverino}, D., {Glover}, S. C.~O., \& {Klessen}, R.~S. 2017, \mnras, 470,
  2791, \dodoi{10.1093/mnras/stx1386}

\bibitem[{{Chiaki} {et~al.}(2017){Chiaki}, {Tominaga}, \& {Nozawa}}]{chiaki17}
{Chiaki}, G., {Tominaga}, N., \& {Nozawa}, T. 2017, \mnras, 472, L115,
  \dodoi{10.1093/mnrasl/slx163}

\bibitem[{{Clark} {et~al.}(2011){Clark}, {Glover}, {Smith}, {Greif}, {Klessen},
  \& {Bromm}}]{Clark11}
{Clark}, P.~C., {Glover}, S.~C.~O., {Smith}, R.~J., {et~al.} 2011, Science,
  331, 1040, \dodoi{10.1126/science.1198027}

\bibitem[{{Dayal} \& {Ferrara}(2018)}]{dayal18}
{Dayal}, P., \& {Ferrara}, A. 2018, \physrep, 780, 1,
  \dodoi{10.1016/j.physrep.2018.10.002}

\bibitem[{{Finkelstein} {et~al.}(2022){Finkelstein}, {Bagley}, {Song},
  {Larson}, {Papovich}, {Dickinson}, {Finkelstein}, {Koekemoer}, {Pirzkal},
  {Somerville}, {Yung}, {Behroozi}, {Ferguson}, {Giavalisco}, {Grogin},
  {Hathi}, {Hutchison}, {Jung}, {Kocevski}, {Kawinwanichakij}, {Rojas-Ruiz},
  {Ryan}, {Snyder}, \& {Tacchella}}]{Finkel22}
{Finkelstein}, S.~L., {Bagley}, M., {Song}, M., {et~al.} 2022, \apj, 928, 52,
  \dodoi{10.3847/1538-4357/ac3aed}

\bibitem[{{Harikane} {et~al.}(2022){Harikane}, {Inoue}, {Mawatari},
  {Hashimoto}, {Yamanaka}, {Fudamoto}, {Matsuo}, {Tamura}, {Dayal}, {Yung},
  {Hutter}, {Pacucci}, {Sugahara}, \& {Koekemoer}}]{Harikane22}
{Harikane}, Y., {Inoue}, A.~K., {Mawatari}, K., {et~al.} 2022, \apj, 929, 1,
  \dodoi{10.3847/1538-4357/ac53a9}

\bibitem[{Harris {et~al.}(2020)Harris, Millman, van~der Walt, Gommers,
  Virtanen, Cournapeau, Wieser, Taylor, Berg, Smith, Kern, Picus, Hoyer, van
  Kerkwijk, Brett, Haldane, del R{'{\i}}o, Wiebe, Peterson,
  G{'{e}}rard-Marchant, Sheppard, Reddy, Weckesser, Abbasi, Gohlke, \&
  Oliphant}]{harris20}
Harris, C.~R., Millman, K.~J., van~der Walt, S.~J., {et~al.} 2020, Nature, 585,
  357, \dodoi{10.1038/s41586-020-2649-2}

\bibitem[{{Hartwig} {et~al.}(2022){Hartwig}, {Magg}, {Chen}, {Tarumi}, {Bromm},
  {Glover}, {Ji}, {Klessen}, {Latif}, {Volonteri}, \& {Yoshida}}]{hartwig22}
{Hartwig}, T., {Magg}, M., {Chen}, L.-H., {et~al.} 2022, arXiv e-prints,
  arXiv:2206.00223.
\newblock \doarXiv{2206.00223}

\bibitem[{{Heger} \& {Woosley}(2002)}]{Heger02}
{Heger}, A., \& {Woosley}, S.~E. 2002, \apj, 567, 532, \dodoi{10.1086/338487}

\bibitem[{Hunter(2007)}]{hunter07}
Hunter, J.~D. 2007, Computing in Science \& Engineering, 9, 90,
  \dodoi{10.1109/MCSE.2007.55}

\bibitem[{{Jaacks} {et~al.}(2019){Jaacks}, {Finkelstein}, \&
  {Bromm}}]{jaacks19}
{Jaacks}, J., {Finkelstein}, S.~L., \& {Bromm}, V. 2019, \mnras, 488, 2202,
  \dodoi{10.1093/mnras/stz1529}

\bibitem[{{Katz} {et~al.}(2022){Katz}, {Kimm}, {Ellis}, {Devriendt}, \&
  {Slyz}}]{katz22}
{Katz}, H., {Kimm}, T., {Ellis}, R.~S., {Devriendt}, J., \& {Slyz}, A. 2022,
  arXiv e-prints, arXiv:2207.04751.
\newblock \doarXiv{2207.04751}

\bibitem[{{Latif} {et~al.}(2022){Latif}, {Whalen}, \& {Khochfar}}]{Latif22}
{Latif}, M.~A., {Whalen}, D., \& {Khochfar}, S. 2022, \apj, 925, 28,
  \dodoi{10.3847/1538-4357/ac3916}

\bibitem[{{Liu} \& {Bromm}(2020)}]{Liu20}
{Liu}, B., \& {Bromm}, V. 2020, \mnras, 497, 2839,
  \dodoi{10.1093/mnras/staa2143}

\bibitem[{{Ma} {et~al.}(2018){Ma}, {Hopkins}, {Garrison-Kimmel},
  {Faucher-Gigu{\`e}re}, {Quataert}, {Boylan-Kolchin}, {Hayward}, {Feldmann},
  \& {Kere{\v{s}}}}]{ma18}
{Ma}, X., {Hopkins}, P.~F., {Garrison-Kimmel}, S., {et~al.} 2018, \mnras, 478,
  1694, \dodoi{10.1093/mnras/sty1024}

\bibitem[{Magg {et~al.}(2022)Magg, Hartwig, Chen, \& Tarumi}]{magg22}
Magg, M., Hartwig, T., Chen, L.-H., \& Tarumi, Y. 2022, Journal of Open Source
  Software, 7, 4417, \dodoi{10.21105/joss.04417}

\bibitem[{{Naidu} {et~al.}(2022){Naidu}, {Oesch}, {van Dokkum}, {Nelson},
  {Suess}, {Whitaker}, {Allen}, {Bezanson}, {Bouwens}, {Brammer}, {Conroy},
  {Illingworth}, {Labbe}, {Leja}, {Leonova}, {Matthee}, {Price}, {Setton},
  {Strait}, {Stefanon}, {Tacchella}, {Toft}, {Weaver}, \& {Weibel}}]{naidu22}
{Naidu}, R.~P., {Oesch}, P.~A., {van Dokkum}, P., {et~al.} 2022, arXiv
  e-prints, arXiv:2207.09434.
\newblock \doarXiv{2207.09434}

\bibitem[{{Oesch} {et~al.}(2016){Oesch}, {Brammer}, {van Dokkum},
  {Illingworth}, {Bouwens}, {Labb{\'e}}, {Franx}, {Momcheva}, {Ashby}, {Fazio},
  {Gonzalez}, {Holden}, {Magee}, {Skelton}, {Smit}, {Spitler}, {Trenti}, \&
  {Willner}}]{Oesch16}
{Oesch}, P.~A., {Brammer}, G., {van Dokkum}, P.~G., {et~al.} 2016, \apj, 819,
  129, \dodoi{10.3847/0004-637X/819/2/129}

\bibitem[{{Omukai} {et~al.}(2005){Omukai}, {Tsuribe}, {Schneider}, \&
  {Ferrara}}]{Omukai05}
{Omukai}, K., {Tsuribe}, T., {Schneider}, R., \& {Ferrara}, A. 2005, \apj, 626,
  627, \dodoi{10.1086/429955}

\bibitem[{{O'Shea} {et~al.}(2015){O'Shea}, {Wise}, {Xu}, \& {Norman}}]{oShea15}
{O'Shea}, B.~W., {Wise}, J.~H., {Xu}, H., \& {Norman}, M.~L. 2015, \apjl, 807,
  L12, \dodoi{10.1088/2041-8205/807/1/L12}

\bibitem[{{Pallottini} {et~al.}(2022){Pallottini}, {Ferrara}, {Gallerani},
  {Behrens}, {Kohandel}, {Carniani}, {Vallini}, {Salvadori}, {Gelli},
  {Sommovigo}, {D'Odorico}, {Di Mascia}, \& {Pizzati}}]{pallottini22}
{Pallottini}, A., {Ferrara}, A., {Gallerani}, S., {et~al.} 2022, \mnras, 513,
  5621, \dodoi{10.1093/mnras/stac1281}

\bibitem[{{Press} \& {Schechter}(1974)}]{PressSchechter}
{Press}, W.~H., \& {Schechter}, P. 1974, \apj, 187, 425, \dodoi{10.1086/152650}

\bibitem[{Price-Whelan {et~al.}(2018)Price-Whelan, Sip{\H{o}}cz, G{\"u}nther,
  Lim, Crawford, Conseil, Shupe, Craig, Dencheva, Ginsburg,
  {et~al.}}]{price2018astropy}
Price-Whelan, A.~M., Sip{\H{o}}cz, B., G{\"u}nther, H., {et~al.} 2018, The
  Astronomical Journal, 156, 123

\bibitem[{{Rigby} {et~al.}(2022)}]{rigby22}
{Rigby}, J., {et~al.} 2022, arXiv e-prints, arXiv:2207.05632.
\newblock \doarXiv{2207.05632}

\bibitem[{{Schaerer} {et~al.}(2022){Schaerer}, {Marques-Chaves}, {Oesch},
  {Naidu}, {Barrufet}, {Izotov}, \& {Guseva}}]{schaerer22}
{Schaerer}, D., {Marques-Chaves}, R., {Oesch}, P., {et~al.} 2022, arXiv
  e-prints, arXiv:2207.10034.
\newblock \doarXiv{2207.10034}

\bibitem[{{Schauer} {et~al.}(2020){Schauer}, {Drory}, \& {Bromm}}]{schauer20}
{Schauer}, A. T.~P., {Drory}, N., \& {Bromm}, V. 2020, \apj, 904, 145,
  \dodoi{10.3847/1538-4357/abbc0b}

\bibitem[{{Schauer} {et~al.}(2021){Schauer}, {Glover}, {Klessen}, \&
  {Clark}}]{schauer21}
{Schauer}, A. T.~P., {Glover}, S. C.~O., {Klessen}, R.~S., \& {Clark}, P. 2021,
  \mnras, 507, 1775, \dodoi{10.1093/mnras/stab1953}

\bibitem[{{Schneider} {et~al.}(2003){Schneider}, {Ferrara}, {Salvaterra},
  {Omukai}, \& {Bromm}}]{Schneider03}
{Schneider}, R., {Ferrara}, A., {Salvaterra}, R., {Omukai}, K., \& {Bromm}, V.
  2003, \nat, 422, 869

\bibitem[{{Skinner} \& {Wise}(2020)}]{skinner20}
{Skinner}, D., \& {Wise}, J.~H. 2020, \mnras, 492, 4386,
  \dodoi{10.1093/mnras/staa139}

\bibitem[{{Stacy} {et~al.}(2016){Stacy}, {Bromm}, \& {Lee}}]{Stacy16}
{Stacy}, A., {Bromm}, V., \& {Lee}, A.~T. 2016, \mnras, 462, 1307,
  \dodoi{10.1093/mnras/stw1728}

\bibitem[{{Sugimura} {et~al.}(2020){Sugimura}, {Matsumoto}, {Hosokawa},
  {Hirano}, \& {Omukai}}]{Sugi20}
{Sugimura}, K., {Matsumoto}, T., {Hosokawa}, T., {Hirano}, S., \& {Omukai}, K.
  2020, \apjl, 892, L14, \dodoi{10.3847/2041-8213/ab7d37}

\bibitem[{Van~Rossum \& Drake(2009)}]{python09}
Van~Rossum, G., \& Drake, F.~L. 2009, Python 3 Reference Manual (Scotts Valley,
  CA: CreateSpace)

\bibitem[{Virtanen {et~al.}(2020)Virtanen, Gommers, Oliphant, Haberland, Reddy,
  Cournapeau, Burovski, Peterson, Weckesser, Bright, {van der Walt}, Brett,
  Wilson, Millman, Mayorov, Nelson, Jones, Kern, Larson, Carey, Polat, Feng,
  Moore, {VanderPlas}, Laxalde, Perktold, Cimrman, Henriksen, Quintero, Harris,
  Archibald, Ribeiro, Pedregosa, {van Mulbregt}, \& {SciPy 1.0
  Contributors}}]{virtanen20}
Virtanen, P., Gommers, R., Oliphant, T.~E., {et~al.} 2020, Nature Methods, 17,
  261, \dodoi{10.1038/s41592-019-0686-2}

\bibitem[{{Warren} {et~al.}(2006){Warren}, {Abazajian}, {Holz}, \&
  {Teodoro}}]{Warren06}
{Warren}, M.~S., {Abazajian}, K., {Holz}, D.~E., \& {Teodoro}, L. 2006, \apj,
  646, 881, \dodoi{10.1086/504962}

\bibitem[{{Windhorst} {et~al.}(2018){Windhorst}, {Timmes}, {Wyithe},
  {Alpaslan}, {Andrews}, {Coe}, {Diego}, {Dijkstra}, {Driver}, {Kelly}, \&
  {Kim}}]{Winhorst18}
{Windhorst}, R.~A., {Timmes}, F.~X., {Wyithe}, J. S.~B., {et~al.} 2018, \apjs,
  234, 41, \dodoi{10.3847/1538-4365/aaa760}

\bibitem[{{Xu} {et~al.}(2016){Xu}, {Wise}, {Norman}, {Ahn}, \& {O'Shea}}]{xu16}
{Xu}, H., {Wise}, J.~H., {Norman}, M.~L., {Ahn}, K., \& {O'Shea}, B.~W. 2016,
  \apj, 833, 84, \dodoi{10.3847/1538-4357/833/1/84}

\bibitem[{{Yan} {et~al.}(2022){Yan}, {Ma}, {Ling}, {Cheng}, {Huang}, \&
  {Zitrin}}]{Yan22}
{Yan}, H., {Ma}, Z., {Ling}, C., {et~al.} 2022, arXiv e-prints,
  arXiv:2207.11558.
\newblock \doarXiv{2207.11558}

\bibitem[{{Yoshida} {et~al.}(2008){Yoshida}, {Omukai}, \&
  {Hernquist}}]{Yoshida08}
{Yoshida}, N., {Omukai}, K., \& {Hernquist}, L. 2008, Science, 321, 669,
  \dodoi{10.1126/science.1160259}

\bibitem[{{Zackrisson} {et~al.}(2011){Zackrisson}, {Rydberg}, {Schaerer},
  {{\"O}stlin}, \& {Tuli}}]{zackrisson11}
{Zackrisson}, E., {Rydberg}, C.-E., {Schaerer}, D., {{\"O}stlin}, G., \&
  {Tuli}, M. 2011, \apj, 740, 13, \dodoi{10.1088/0004-637X/740/1/13}

\bibitem[{{Zackrisson} {et~al.}(2017){Zackrisson}, {Binggeli}, {Finlator},
  {Gnedin}, {Paardekooper}, {Shimizu}, {Inoue}, {Jensen}, {Micheva},
  {Khochfar}, \& {Dalla Vecchia}}]{zackrisson17}
{Zackrisson}, E., {Binggeli}, C., {Finlator}, K., {et~al.} 2017, \apj, 836, 78,
  \dodoi{10.3847/1538-4357/836/1/78}

\end{thebibliography}
\bibliographystyle{aasjournal}



\end{document}